\documentclass[10pt, conference, compsocconf]{IEEEtran}

\usepackage{times}
\usepackage{latexsym}
\usepackage[pdftex]{graphicx}
\usepackage{subfig}
\usepackage{url}
\usepackage{hyperref}
\usepackage{amsmath}
\usepackage[]{algorithm2e}
\usepackage{float}
\usepackage{textcomp}

\begin{document}

\title{Technical Health Check For Cloud Service Providers}

\author{
    \IEEEauthorblockN{Muhammed Fatih Bulut\IEEEauthorrefmark{1}, Hongtan Sun\IEEEauthorrefmark{1}, Pritpal Arora\IEEEauthorrefmark{2}, Maja Vukovic\IEEEauthorrefmark{1}, Klaus Koenig\IEEEauthorrefmark{2}, Jonathan Young\IEEEauthorrefmark{2}}
    \IEEEauthorblockA{\IEEEauthorrefmark{1}IBM T.J. Watson Research Center
    \\\{mfbulut@us.ibm.com, hongtan.sun@ibm.com, maja@us.ibm.com\}}
    \IEEEauthorblockA{\IEEEauthorrefmark{2}IBM Services
    \\\{pritpal.arora@in.ibm.com, kkoening@de.ibm.com, jonathan.r.young@uk.ibm.com\}}
}

\maketitle

\IEEEpeerreviewmaketitle

\begin{abstract}

Understanding the overall health of an IT Infrastructure is a key part of IT Service Management. Traditional approach to perform technical health check is by visiting customer's physical site and rigorously examining the IT infrastructure with Subject Matter Experts. Alternatively, periodic surveys are sent to Technical Architects who are responsible for the managed IT infrastructure. In essence, both site visits and surveys suffer from reactive nature, and subjective assessment. In this paper, we present technical health check for cloud providers, that monitors, assesses operational data and depicts the current health of an IT infrastructure in real time. We also discuss challenges and opportunities of technical health check in Hybrid Cloud Environment.

\end{abstract}


\section{Introduction}

Enterprises are increasingly outsourcing their IT management to focus on their core business. IT service providers are tasked with ensuring that the customer's IT infrastructure is healthy, secure and operating seamlessly by managing the core elements of IT environments such as, servers, storage and middleware. Moreover, service providers also need to make sure that  industry specific regulations are enforced to avoid audit failures and penalties.

Technical Health Check (THC) is a service management practice, which traditionally entails visiting customer's physical site and work with customer's Subject Matter Experts (SMEs) to examine all of the IT infrastructure in order to understand problems and resolve them. The process of visiting the customer's physical site can be either in a reactive way, by responding to a major incident reported by the customer, or a proactive approach in which periodic visits are made and overall health of different IT components are rigorously examined. 

Despite the effectiveness of working with customer's SMEs face-to-face, site visits are costly and more importantly not scalable. Many SMEs (both from Service Provider and Customer) need to be present and work together to find out issues and resolve them. Therefore, usually site visits are limited to once or twice a year. As a natural outcome, site visits neither prevent all the incidents taking place and customers still need to rely on a reactive approach where incidents happen and SMEs visit customer's physical site to fix the issues; nor can help optimize the current setup.

Alternative to site visits is to conduct periodic surveys with the Technical Architects (TAs), who are responsible for the managed IT environment for the customer. The idea with conducting a survey is to ask several questions to the TA and expect a fair and accurate response on the health of the IT environment. The downside of this approach is that since TA is the main person responsible for the IT environment, sometimes biased and optimistic answers are given which make it hard to make use of the survey data to understand the overall health. Number of questions, and the variety of expertise needed to answer these questions are also other downsides of the survey based approach, and hence falls short of solving issues that the customer faces. 

To overcome these challenges, this paper presents Technical Health Check for cloud service providers, which automates the process by utilizing operational data collected from various components of the IT infrastructure and provides insights through monitoring, analyzing and quantifying operational data to various health controls, called Key Performance Indicators (KPIs) --- derived from COBIT\textregistered{} \footnote{COBIT\textregistered{} is a registered trademark of the Information Systems Audit and Control Foundation. \url{http://www.isaca.org/cobit/pages/default.aspx}} framework. Our framework enables IT service providers to quantitatively measure the technical health check of the managed IT infrastructure. The contributions of the paper are as follows:

\begin{itemize}

\item We presented the building blocks of Technical Health Check (THC) for cloud providers. THC is a cloud-based service for IT service providers which utilizes operational data collected from various IT components, as well as from Change, Incident and Event Management Systems and make sense of the collected data to depict the health of an IT environment. 

\item We presented an approach to derive Key Performance Indicators (KPIs) quantitatively. KPIs reflect different critical health controls using industry-standard COBIT\textregistered{} framework. 

\item We presented various use cases of KPIs, namely Generating Heatmaps to reflect the account's health status, KPI prediction to estimate future critical degradations, and finally KPI correlation analysis to unveil hidden correlations among KPIs.

\item Finally, while we focus on the traditional IT service management; we provide a discussion on challenges and opportunities that Hybrid Cloud brings from the perspective of Cloud Service providers.

\end{itemize}

The rest of the paper is organized as follows: In Section \ref{sec:related}, we discuss the related work. In Section \ref{sec:thcaas}, we discuss the building blocks of our THC framework. In Section \ref{sec:use_cases}, we layout various use cases of KPIs in our THC framework. Next, in Section \ref{sec:cloud_native}, we discuss how Hybrid and Multi Cloud trends can help better understand the overall health of an IT infrastructure as well as the additional challenges that they bring. Finally we conclude with Section \ref{sec:conclusion}.

\section{Related Work}
\label{sec:related}

Correctly assessing Technical Health Check (THC) of a managed IT environment is a quintessential part of service provider's duty. There are variety of tools and extensive literature that are related to the technical health check of a particular component of an IT environment such as application \footnote{\url{https://newrelic.com}}, \footnote{\url{https://www.datadoghq.com}} \cite{rak2011cloud, alhamazani2014clams}, storage \footnote{\url{https://www.appdynamics.com}} and network \footnote{\url{https://www.solarwinds.com}}. In addition, there are many tools out there to monitor the resources (such as CPU, Memory) in Cloud \cite{aceto2013cloud, aws_cloud_watch}. However there are two main differences of our work to previous work. First, our THC framework integrates raw Operational Data and aggregate them as Operational Metrics (OMs) and transform OMs to the Key Performance Indicators (KPIs) that provides a holistic view of the IT environment. Second, our framework integrates Subject Matter Experts (SMEs) input on KPI score calculation, and allow them to specify and dynamically change which components or health controls weights more among others.

In a related domain, Incident and ticket analysis is an active research area within the community. In \cite{diao2009rule} Diao et al. proposes to use crowds' ability to define classification rules for tickets where there are insufficient classified data. In \cite{DBLP:journals/tnsm/ZengZLSG17} authors propose multi-label classification system for IT tickets utilizing the category hierarchy. In \cite{DBLP:conf/kdd/ZhouXBWZLXLSG17}, Zhou et al. developed a ticket analysis framework which recommends a solution for a given ticket using the previous history of resolutions. 

Change Management is another source of operational data that is utilized by our THC framework. In \cite{kadar2011automatic}, authors propose to auto-classify the incoming change requests to better understand the impact by aggregated information associated with the change, for example reasons for past failures or best implementation practices. In \cite{DBLP:conf/icsoc/KaliaXBVA17}, Kalia et al. suggest a framework that auto-classify change requests to the classes in catalog using machine learning. In another paper, Guven et al. \cite{DBLP:conf/noms/GuvenMSP16}, develop a solution to build casuality between change and incident, which can be used to understand change management maturity.

Assessing the Security \& Compliance in Cloud is an up-most importance to our framework. Organizations like CIS and NIST provide guidance on how compliance should be done in managed IT infrastructure via technical specifications and vulnerability reporting and scoring \footnote{\url{https://www.cisecurity.org}}, \footnote{\url{https://www.nist.gov}}. In \cite{DBLP:conf/services/FileppAHVAZ18} explain a solution that enforce continuous compliance in a managed IT infrastructure. Besides, there are many work in literature that discuss how security and compliance can be achieved in Cloud \cite{Krutz:2010:CSC:1869722, hendre2015semantic, bhensook2012assessment}.

\section{Technical Health Check Architecture}
\label{sec:thcaas}

\begin{figure*}[ht]
  \begin{center}
    \includegraphics[width=1.95\columnwidth,
      keepaspectratio]{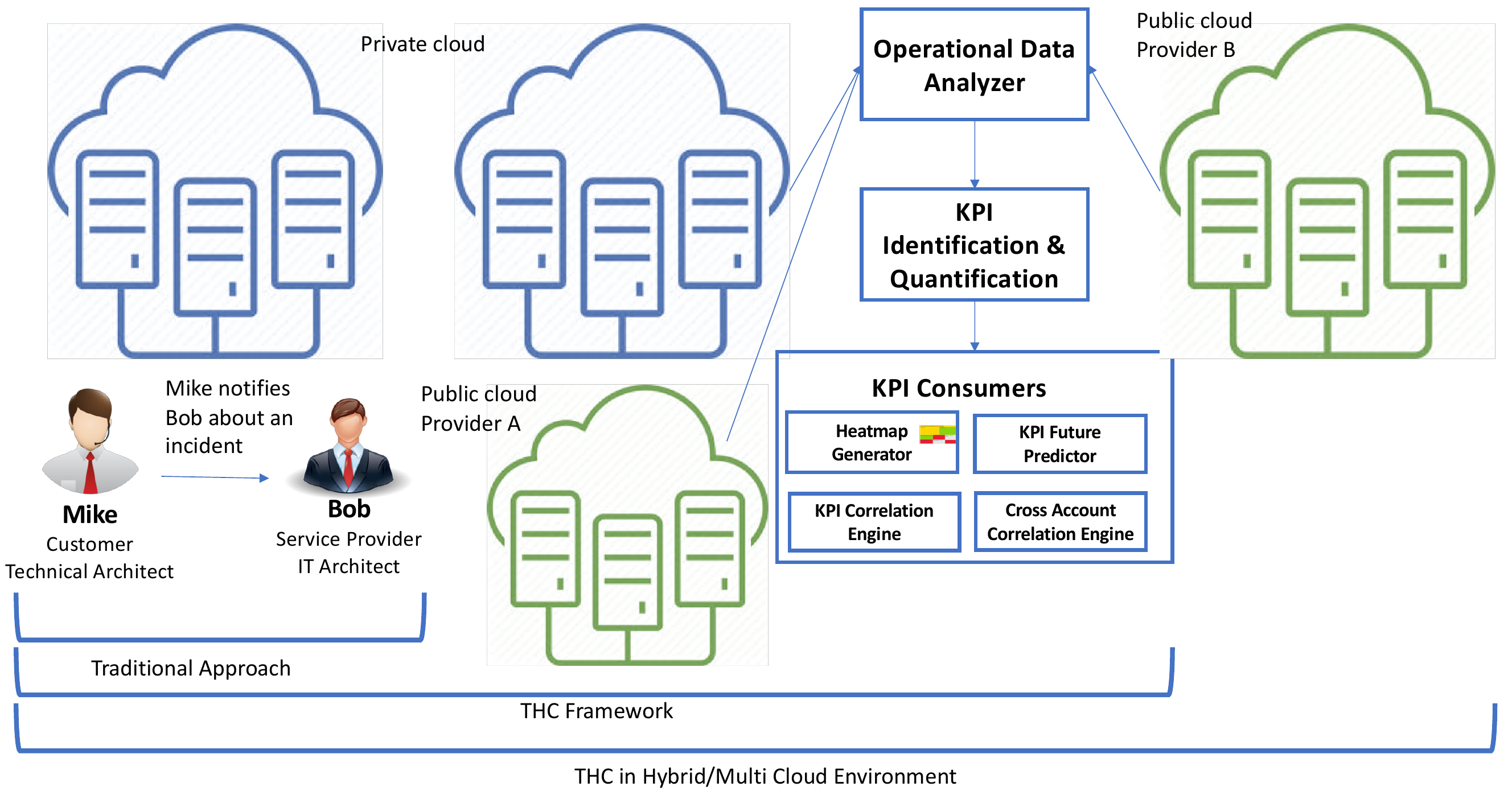}
    \caption{The evolution of Technical Health Check for managed IT infrastructure.}
    \label{fig:arch}
  \end{center}
\end{figure*}

Our Technical Health Check (THC) framework consists of three core components as shown in Figure \ref{fig:arch}: Operational Data Analyzer Module, KPI Identification \& Quantification Module and KPI Consumers Module which consists of several functional use cases of KPIs and will be explained in Section \ref{sec:use_cases}. In the following subsections, we will delve into the details of the first two components of our THC framework.

\subsection{Operational Data Analyzer}

Operational data is an invaluable asset to understand the overall health of an IT infrastructure. Our framework mainly focuses on the operational data that is coming from Asset Management, Change Management, Event Management, Incident/Ticket Management and Security \& Compliance Management. 

\emph{Asset Management} provides information about the age of the physical infrastructures such as servers and switches. As a device gets older, it is getting closer to End-of-Life (EoL) or End-of-Service (EoS). In that case, it is likely that these devices will fail more, and when they fail, the official technical support would be potentially unavailable. So the age of the assets is closely related to the performance of the whole IT infrastructure system and is an important indicator of its overall health. 

\emph{Change Management} allows service providers to keep track of the changes that have been applied to the various components of the IT infrastructure such as OSes, Middlewares and Applications. For example if a new memory needs to be attached to a server, it needs to be recorded on Change Management system first. As another example, if the operating system needs to be updated for a virtual machine, a change request also has to be submitted to the Change Management system and the updating process will be kept in the change request record. 

\emph{Event Management} provides a way to monitor various defined events on the service provider's radar. For example if a storage utilization of a server exceeds a certain threshold, an event may be fired and as a consequence different actions can be taken such as adding more capacity to the server. Such record keeping helps both service providers to optimize its management operations, as well as makes process/people accountable for audits.

\emph{Incident \& Ticket Management} provides a way to monitor the incidents that occur on the infrastructure as reported by the engineers or specialists. For example, a ticket may be opened for an application failure. It may turn out that the application failure is because of a recent change that has been made to the same server on its firewall rules. Therefore, utilizing the linkage between different management services, in this case Incident and Change Management, enable service providers to find the root cause of the problems. However, understanding the tickets comes with its own challenges. Most of the tickets are written in plain text in natural language and hence needs to be understood, i.e proper natural language processing and machine learning techniques (such as classification) needs to be applied to correctly understand the problem reported in the ticket. 

Finally, \emph{Security \& Compliance Management} provides immense value as to evaluate the security and the compliance of the infrastructure in regards to various regulations that are asked to be enforced by the customer. 

In a nutshell, Operational Metric (OM) is an aggregation of raw operational data. To illustrate the advantage of having aggregated measurements like OMs over the raw operational data, consider when updating a library on a Virtual Machine. Usually every update needs to be taken care of with care, meaning pre and post steps need to be followed in order not to disrupt the other services or applications running on the same VM. For example, in order to update a database, one also needs to make sure that application which uses the database needs to be taken care of first (either by shutting down or taking some other measurements to make sure that nothing gets broke). To perform such series of actions, usually a change window (of limited time) is dedicated and customer is notified beforehand. It could happen that the update to the library is not completed within the change window, in which case the change request is marked as incomplete or backlogged. If an account has a lot of such backlogged changes, it usually indicates an issue in the Change Management System. This shows that aggregated Operational Metrics (OM), such as \emph{\% Backlogged Change Requests} from the raw operational data provides insights of the overall health condition for the Change Management Operations. 

\begin{table}[t]
\centering
\begin{tabular}{l l l l l}
\hline
 Description & Low & Up & Scale & Dir.  \\ \hline
 \% of Failed Changes & 0 & 100 & Configured & Min  \\
 \% of unauthorized changes & 0 & 100	& configured & Min \\
 \% of emergency changes & 0 & 100 & configured & Min \\
 \% of changes resulting in incidents & 0 & 100 & configured & Min \\
 \% of rescheduled changes & 0 & 100 & configured & Min \\
 \% Change Backlog   - Total & 0 & 100 & configured & Min \\
 Failure rate for server-related changes & - & - & Captured & Min \\
 \% servers with auto patch management & 0 & 100 & configured & Max \\ 
 \% Servers enabled for monitoring & 0 & 100 & configured & Max \\ 
 Security Health Check Risk Score & 0 & 10 & configured & Min \\ 
\end{tabular}
\caption{Examples of Operational Metrics}
\label{tab:change_mgmt} 
\end{table}

Table \ref{tab:change_mgmt} gives examples of OMs. As can be seen from the Table, for each OM, we have a description, Lower and Upper bounds, Scale Type and Direction (to understand whether to minimize or maximize the the metric for a better performance). In order to quantify the OMs in a standard way across the metrics, we need a \emph{normalization process}. Let us now delve into how normalization process works. 

For each OM, we get a performance score which can be measured by different units. In order to have a valid comparison, all elements of performance values across OMs need to be transferred into the same unit. Thus the approach is the following:

\begin{itemize}

\item The subject matter expert determines the classification type of OM. There are two options of classification types: a) Maximization Type: OM is to be maximized, meaning highest value is the best value, b) Minimization Type: OM is to be minimized, meaning lowest value is the best value. For example, an OM such as \emph{Percentage of Failed Changes} should be minimized and hence lower the value is the better it is.

\item Sometimes upper and lower values for OM needs to be defined based on other customer's overall performance. Scale type of `Captured` allows us to see if bounds are defined dynamically. As can be seen from Table \ref{tab:change_mgmt}, \emph{Failure rate for server-related changes} are defined as Captured and their lower and upper bounds are empty to indicate dynamic changes for the values.

\item If the Scale type is `Configured`, for each OM of maximization/minimization type, the Subject Matter Expert (SME) defines two values: lower boundary for the worst performance and upper limit for the best performance (or the vice a versa in case of minimization).

\item In general, the acceptance level values (lower and upper bounds) can be different for the different OMs. Dependent of the OM's maximization or minimization type we define a norm function for the OM's performance values. There are different techniques available, such as Min-Max Normalization, Vector Normalization, Linear Sum-based Normalization and Logarithmic Normalization. In our case, we use \emph{Min-Max Normalization} with a range $[0, 10]$, so our formula becomes as shown in Equation \ref{eq:metric_val}. 

\item To account for the case where Scale Type is minimization, we subtract the value in Equation \ref{eq:metric_val} from max-range value, which is $10$ but can be easily configurable to other values.

\end{itemize}

\begin{equation}
  o_i = \frac{Actual Value - Lower Bound}{Upper Bound - Lower Bound} \times 10
\label{eq:metric_val}
\end{equation}

By this procedure, we achieve two goals: All OMs performance values are transformed to the unit interval $[0, 10]$, and all the OMs of type minimization are converted into the criteria of maximization type thus can be handled together in subsequent steps as equal citizens.

\subsection{KPI Identification \& Quantification}

Operational Metrics (OMs) provide an aggregated measurement for a particular unit of function in the IT infrastructure. OMs are derived from many sources which are available to IT service providers. As a result, usually there are multiple metrics to understand the overall health of the particular health control. For example, for change management maturity as can be seen from Table \ref{tab:change_mgmt}, we have multiple OMs to account for Change Management, and not all of them necessarily maps to one Key Performance Indicator. Hence, to better understand the overall health of the infrastructure, we need higher level metrics than OMs (which captures finer granularity functionalities). We call these higher level metrics, \emph{Key Performance Indicators (KPIs)}. KPIs capture the essence of the particular IT Environment's management performance. 

To define KPIs we rely on an industry standard framework called COBIT\textregistered{}. In a nutshell, COBIT\textregistered{} provides a control framework for IT governance, aligns business goals with enterprise IT architecture. COBIT\textregistered{} specifies key indicators and measurements to control in the IT management system, hence guides us the performance indicators in the service quality control taking the business goals into the account. By using COBIT\textregistered{}, we rely on an industry standard framework to be able to account the health of the IT environment. Let us now explain how mapping from OMs to KPIs is done in our framework. 

KPIs captures the health of a critical IT component for a Service Provider. Using KPIs, service providers, particularly Technical Architects, can quantitatively assess the health of particular crucial component of the IT environment and reason about potential technical issues, and impact to the business. Below we briefly describe our methodology to quantify the KPIs. 

\begin{table}[t]
\centering
\begin{tabular}{l l l}
\hline
	OM & Weight & Function  \\ \hline
	Database Space Issue(N) tickets & 0.25 & Linear \\
	Database Handler tickets & 0.25 & Linear \\
	DB2 Instance Down(A) tickets & 0.25 & Linear \\
	Database Job Warning(N) tickets & 0.25 & Linear \\
\end{tabular}
\caption{Mapping of Operational Metrics to KPI (Database Resiliency Management)}
\label{tab:mapping}
\end{table}

Given OMs, we need to combine relevant OMs and come up with a score. Usually OM to KPI mapping is many-to-many mapping. Note that different OMs can have different impact on the KPI score and therefore it needs to be accounted in the formula. Hence we need a \emph{weighing process}. Thus the approach is the following:

\begin{itemize}
\item First SME defines the importance of each OM contributing to a KPI, as shown in Table \ref{tab:mapping}. However, different approaches such as machine learning to learn the weights can also be applied and we leave it as a future work. 

\item Second we need to combine the different OMs. Currently in our framework, we use a linear weighted combination. However more complex functions/formulas can also be explored (such as Quadratic function etc). More specifically, given OMs $o_1, o_2, ..., o_n$ with corresponding weights as $w_1, w_2, ..., w_n$, we calculate the KPI score as shown in Equation \ref{eq:kpi}, using a scale from $[1, 5]$.
\end{itemize}

\begin{equation}
	kpi_j = 1 + 0.4 \times \sum_{i=0}^{n} o_i \times w_i
\label{eq:kpi}
\end{equation}

\begin{algorithm}
 \KwData{Actual Values For OMs=$[a_0, a_1, ..., a_{n-1}]$, Upper Bounds for OMs=$[u_0, u_1, ..., u_{n-1}]$, Lower Bounds for OMs=$[l_0, l_1, ..., l_{n-1}]$, Scale Type for OMs=$[s_0, s_1, ..., s_{n-1}]$}
 \KwResult{KPI values=$[kpi_0, kpi_1, ..., kpi_{m-1}]$}
 $i \gets 0$\;
 \While{$i < n$}{
  $o_i \gets \frac{a_i - l_i}{u_i - l_i} \times 10 $\\
  \If{$s_i$ $=$ 'Minimization'}{
  	$o_i \gets 10 - o_i$
  }
  $i \gets i + 1$
 }
 $j \gets 0$\;
 \While{$j < m$}{
	$[o_0, o_1, ..., o_{k-1}], [w_0, w_1, ..., w_{k-1}] \gets getAllMappedOMs(kpi_j)$\;
	$i \gets 0$\;
	$sum \gets 0$\;
	\While{$i < k$} {
		$sum \gets sum + w_i \times o_i$\\
		$i \gets i + 1$
	}
	$kpi_j \gets 1 + 0.4 \times sum$\\
	$j \gets j + 1$
}
 \caption{Calculation of KPI Values}
\label{alg:kpi_calc}
\end{algorithm}

Algorithm \ref{alg:kpi_calc} shows the overall calculation process for both OM and KPI values. In overall, the advantage of the method proposed above for multi-criteria (multiple OMs) scoring is the fact that we can deal with different scale types of OM, whether it is a maximization type or a minimization type. Furthermore, even if we have different scales and units of OM, the normalization process can convert all OM into the same unit base. With a simple pairwise evaluation of the most dominant OM against the remaining criteria, the method allows the qualitative input from a subject matter expert’s point of view. With this approach, the expert’s input is easy to understand and can be effectively handled. Likewise, the underlying qualitative model by importance level comparisons can be easily adjusted by machine learning and is suitable for further analysis.

\section{KPIs Use Cases}
\label{sec:use_cases}

KPIs encompass various OMs and quantitatively reflects the health of a particular component of IT environment. In our THC framework, there are various consumers of KPIs. Among others, below we explain four of these consumers and shared their respective experimental results.

\subsection{IT Environment Heatmap}

\begin{figure}[ht]
  \begin{center}
    \includegraphics[width=1\columnwidth,
      keepaspectratio]{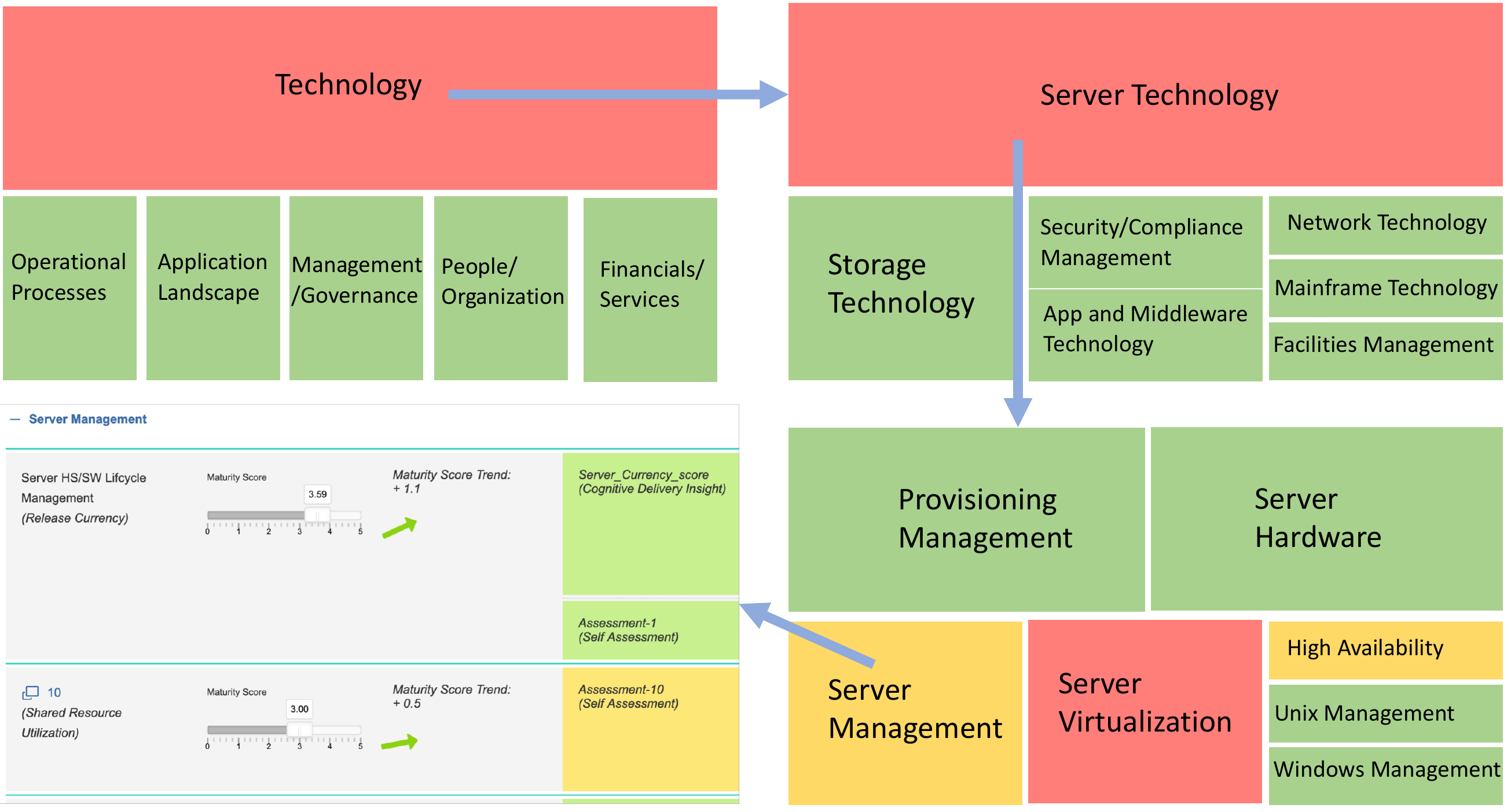}
    \caption{Generated heatmap as drilled down from left to right and top to bottom.}
    \label{fig:heatmap}
  \end{center}
\end{figure}

Once KPIs are calculated, Technical architects can use KPI scores to better understand the health of the managed IT environment. To provide an intuitive expression of the KPI scores, we visualize them on a heatmap as similar to shown in Figure \ref{fig:heatmap}. The figure shows the drill down feature of the heatmap, starting from top left in clockwise. The color scheme is coded in three colors: \emph{Green} indicates a good performance (KPI score of $[4,5]$), \emph{Yellow} indicates a caution (KPI score of $[2,4)$) and \emph{Red} indicates a problem (KPI score of $[1,2)$) when immediate action is needed to investigate the IT component and fix any issues. The heatmap starts from a high level topic, such as Technology, Operational Processes etc., as shown in Figure \ref{fig:heatmap} and can be drilled down until the actual KPI reached along with associated OMs. Trend is also shown to indicate the change from the last captured value.

\subsection{Benchmarking Among Customers}

\begin{figure*}[ht]
  \begin{center}
    \includegraphics[width=1.9\columnwidth,
      keepaspectratio]{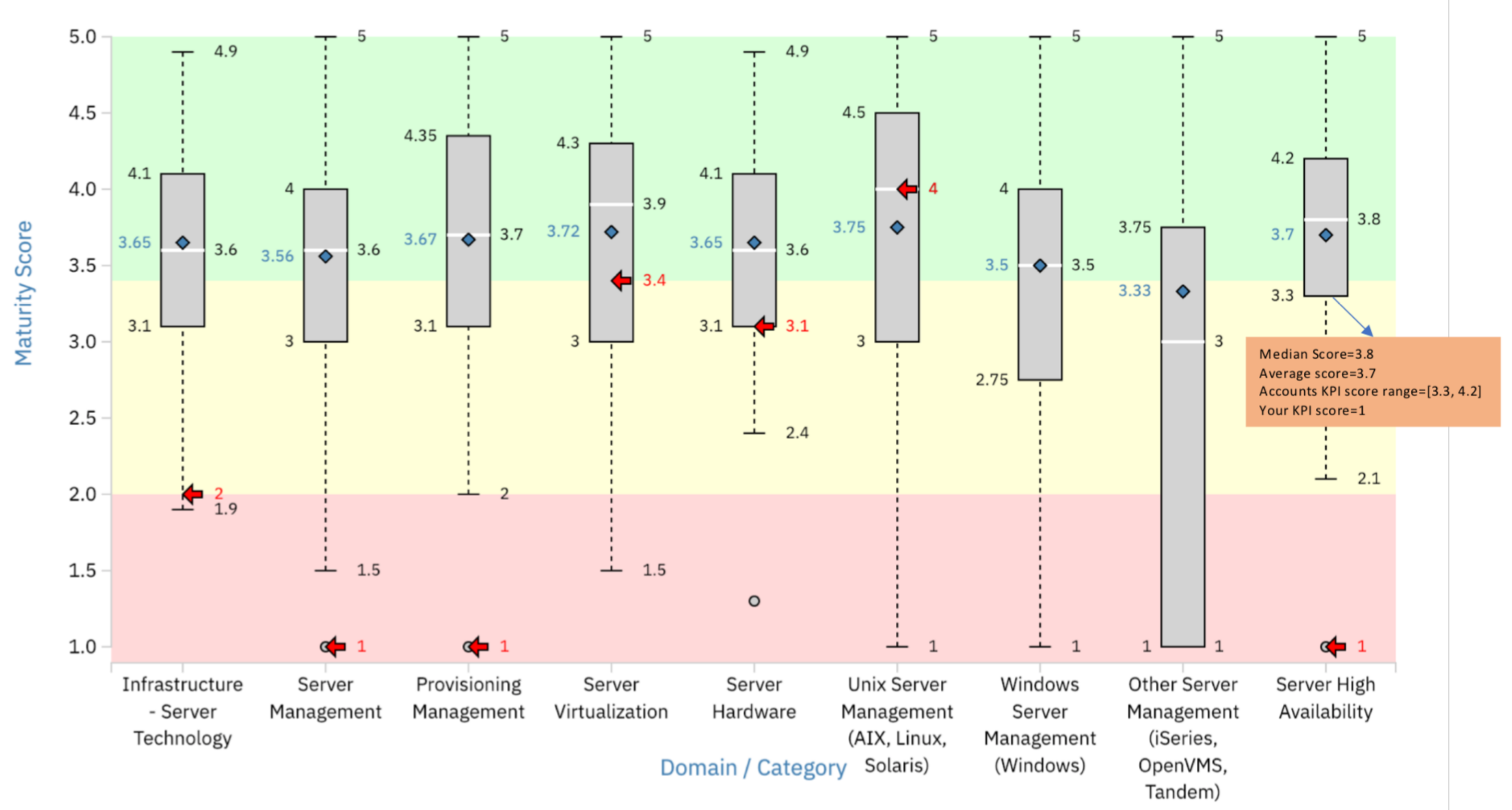}
    \caption{Graph that shows where a customer stand among others in Server Technology Domain.}
    \label{fig:cross_acc}
  \end{center}
\end{figure*}

Another unique features of our THC framework is to be able to benchmark an account among other accounts. Figure \ref{fig:cross_acc} shows an example of benchmarking for an account for Server Domain. As shown in the Figure, overall statistics of all the other accounts (min., max. and median scores) are shown to help Technical Architects to understand where her account stands among others. For example, for \emph{Server High Availability}, account's KPI score is $1$, on the other hand, median score for all the other accounts are $3.8$ in a range from $3.3$ to $4.2$, which clearly indicates a problem in this particular health control for this particular account.

\subsection{KPI Score Prediction}

\begin{figure}[ht]
  \begin{center}
    \includegraphics[width=1\columnwidth,
      keepaspectratio]{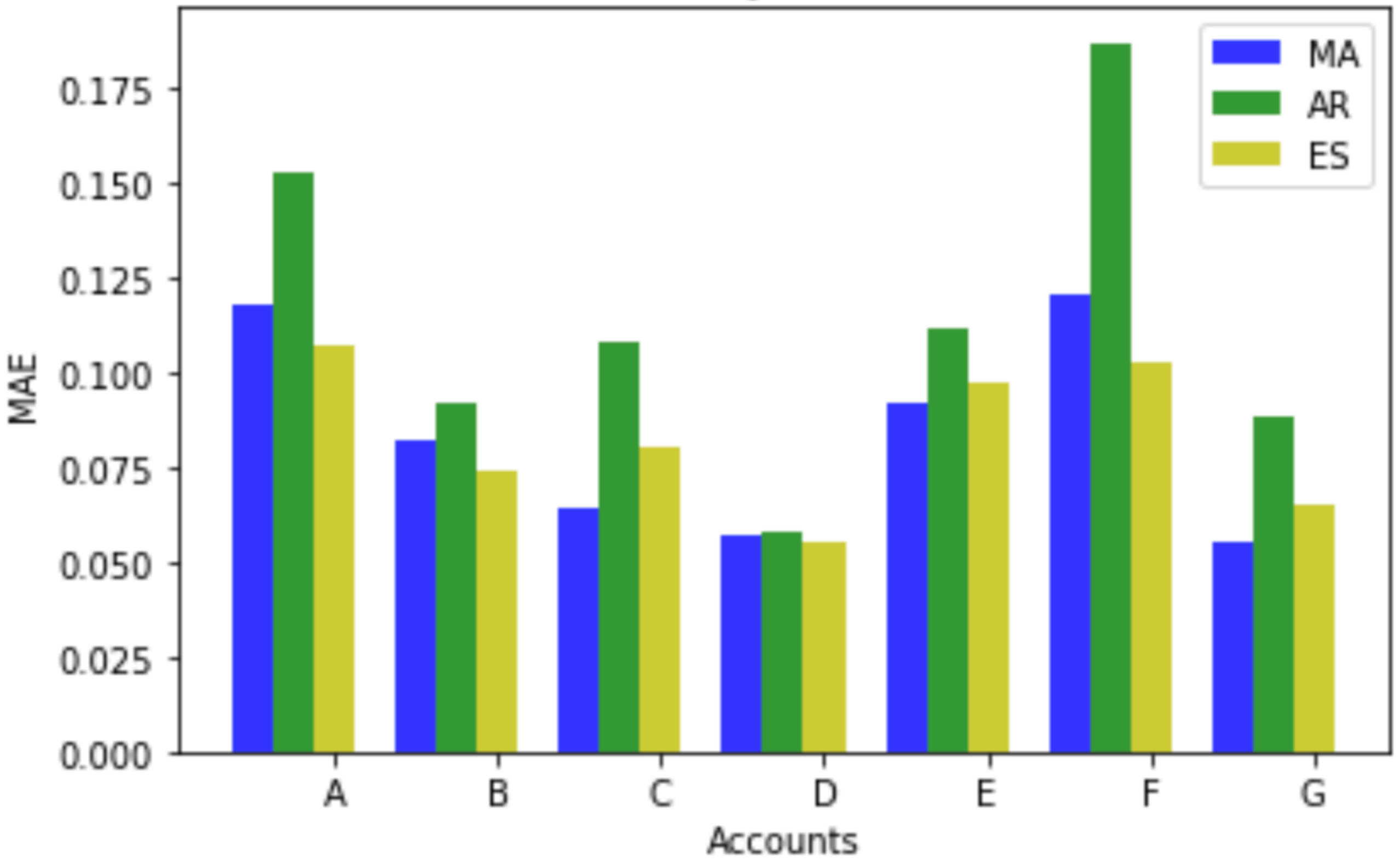}
    \caption{Mean Absolute Errors for Different Accounts. MA: Moving Average, AR: Autoregression, ES: Exponential Smoothing}
    \label{fig:mae_accounts}
  \end{center}
\end{figure}

Though Technical Architects can use KPI scores to understand the overall health of particular health controls, it is also useful to know what will be the score in the near future. With that information available to Technical Architects, they can make more informed choices and act pro-actively. In this part, we present an experiment that we performed over $6$ select KPIs over 12-months period for $7$ accounts to show how affectively we can predict the next month's value. Due to various reasons such as data feed problem, in some months, we were not be able to calculate the KPI values (we have only $14$ such missing values out of $504$ KPI scores). For those missing months, we simply replaced the missing values with the average of their immediate neighbors. 

Our methodology for evaluation is as follows: Starting from 8th month, we use the previous months' data to predict next month's value. Hence, to predict the KPI score for 11th month, we use the last 10 months of scores as training, and calculate the Mean Absolute Error (MEA) as the measurement. We take the average of MAEs (from month $9$ to $12$) and set as the calculated metric for that particular KPI for an account. To predict next month's score, we utilize three widely used time series forecasting methods, namely Autoregression (AR), Moving Average (MA) and Exponential Smoothing (ES). Figure \ref{fig:mae_accounts} shows the average MAE for different accounts. Across all accounts overall MAEs are: MA $0.084$, ES: $0.083$ and AR $0.113$. From the results, it is understood that MA, ES performs similarly and better than AR. Keeping in mind that KPI scores ranges from $[1, 5]$ MAE less than $0.1$ can provide a pretty good estimate of what is going to be the score of a KPI for the next month and can be helpful for Technical Architects.

\subsection{KPI Correlations}

Since there are multiple KPIs covering different aspects of IT service management system, it is important to look at the correlation of one KPI to the other KPIs. This can be quite useful in a way that, correlated KPIs can account to full problem that is observed on the IT environment. As an example, application failures may be a result of a poor capacity management, and if this correlation is observed, we can arrive to a solution quicker than only focusing on application failures. 

In order to calculate KPI correlations, we look at the historical KPI scores. We use 6-months of KPI scores and use the coefficient of correlation formula to calculate correlation between different KPIs. In particular, if the values of KPI1 in the last 6 months are $k_1, k_2, ..., k_6$ and the values for KPI2 in the last 6 months are $p_1, p_2, ..., p_6$, we use the predicted value from linear regression to help find the coefficient of correlation, i.e. $f_i$ for $i = 1, 2, ... , 6$ given in \ref{eq:lin} below.

\begin{equation}
	f_i = k_i\frac{\sum_{j = 1}^{6} k_j \times p_j}{\sum_{j = 1}^{6}k_j^2}.
\label{eq:lin}
\end{equation}

Then the correlation score between KPI1 and KPI2 is in \ref{eq:cor} below:

\begin{equation}
	cor(KPI1, KPI2) = 1 - \frac{\sum_{i = 1}^{6}(k_i - f_i)^2}{\sum_{i = 1}^{6}(k_i - \bar{k})^2},
\label{eq:cor}
\end{equation}

where $\bar{k} = \frac{1}{n} \sum_{i = 1}^{6} k_i$ is the average value of KPI1 over the last 6 months. In the practical implementation, the SciPy package is used to compute the r-square scores between pairwise KPIs.

When the correlation score is above 0.5, empirically it indicates that the two KPIs are strongly related. Identifying strongly related KPIs could help demonstrate hidden relationships between different domains. Table \ref{tab:kpicorr} shows four strongly correlated KPIs we found in one of the experiments. The KPI {\bf Server Monitoring Ratio} and {\bf Server Capacity Events Ratio} measures the percent of server that is being automatically monitored and the percent of CPU/Memory related events in the system, which are both IT operation process related KPIs. While {\bf Business Outage} measures the number of business outages that happened in the past 6 months, which is a business performance related KPI. The KPI scores show that all the three KPIs are strongly correlated with the KPI {\bf DPP}, which refers to {\em Defect Prevention Process}. It indicates that the server monitoring and capacity monitoring have a great impact on the business outage. By discussing with Subject Matter Experts (SMEs), we learned that from their experience, if the accounts' servers and their capacity are better monitored, their risks of having severe business outages are much better controlled.

\begin{table}
\centering
\begin{tabular}{ccc}
\hline
KPI1 & KPI2 & Correlation Score \\
\hline
DPP & Server Monitoring Ratio & $ 0.524$ \\
DPP & Server Capacity Events Ratio & $0.625$ \\
DPP & Business Outage & $0.538$ \\
\end{tabular}
\caption{Strongly correlated KPIs for a demo account}
\label{tab:kpicorr}
\end{table}

\section{Technical Health Check in Hybrid/Multi Cloud Environment}
\label{sec:cloud_native}

With companies are increasingly moving to the public cloud, this brings in new challenges for IT service providers. Additionally, according to the recent statistics, 85\% of the companies are already using more than one cloud environments, and the number is expected to increase to 98\% in 3 years \cite{hybrid_stats}. As such, these trends make service provider's job challenging in two ways.

First, there are variety of companies that provide public cloud services including Amazon, Google, IBM and Microsoft. Each one of these public cloud vendors provides different set of services for their customers. Besides, these services can change rapidly and new services are constantly being added. Both of these factors make it hard for service providers to be able to find Subject Matter Experts (SMEs) who fully understand the services to provide the required services for their customers. Meanwhile, failure to address user's concern in a timely manner would undermine the provider's goodwill.

Second, companies have different motivations to move to the public cloud. Some to reduce the cost, some to take advantage of a new set of managed services provided by the cloud vendors such as the ones related to Artificial Intelligence. At the same time, many companies (especially big ones), have already made a lot of investments over the decades to build up their own datacenters (again for various reasons, such as privacy and security), so they are usually reluctant to move away from their own private datacenter altogether. As a result, increasingly, many companies end up in an IT environment where applications run in hybrid cloud environment. Microservices also support this trend, and it is not very uncommon to see some microservices of an application are running in public cloud and some are running in private cloud. As a result, IT service providers increasingly faces to manage hybrid workloads which are obviously harder than managing IT service provided by one vendor.

On the other side, moving to the public cloud comes with its own benefit and opportunities. First service providers now be able to help customers to move different cloud providers where customers sees more appropriate. In a way, this would help prevent lock-down in a specific cloud vendor and take advantage of the price and service varieties of different cloud providers. Assuming, IT service providers have enough SMEs to support this journey, companies can increasingly would like to take advantage of it. Second, service providers can now offer their customers more choices, and also be able to go along with the customer's strategy than enforcing their own strategy in terms of which provider or service to use. 

Given the above challenges and opportunities, next we discuss how our framework can adapt itself to work in different cloud architectures, such as Infrastructure as a Service (IaaS) and Platform as a Service (PaaS).

\subsection{THC in IaaS}
In IaaS environment, some of the metrics that our THC framework makes use of today is not appropriate. For example, server (physical) currency (age) now becomes responsibility of the public cloud vendor and our framework may not need to worry about. At the same time we still need to keep track of what changes applied to servers, so proper Change Management needs to be in place. Also necessary tools need to be in place to monitor the events and logs in the infrastructure. In some cases, cloud vendors already provides such tools, therefore proper integration of such tools to our framework is necessary to capture the full picture. Same is valid for Security \& Compliance, service providers still need to provide the Security and Compliance requirements based on the customer's need.

\subsection{THC in PaaS}

In PaaS environment, for example in Cloud Foundry \footnote{\url{https://www.cloudfoundry.org}}, both the virtual machines (servers) and runtimes (such as Java, Python etc.) are provided and managed by the cloud providers. In this case, we mostly need to worry about the application. Our framework needs to make sure that application development follows the DevOps rules and secure and compliant applications are pushed to the cloud providers. An opportunity emerges for our framework to provide advisory capabilities through monitoring of an applications and their performance on possible rearchitecting or refactoring based on the operational data and insights. Traditional processes such as, change management need to be adapted to the new operating model. With DevOps the scale of changes increases, as well as the dependency across microservices. These new models of operation will drive and capture the KPIs that are meaningful in new hybrid Cloud environment.

\section{Conclusion}
\label{sec:conclusion}

In this paper, we present Technical Health Check for cloud service providers. Our THC framework makes use of raw operational data as Operational Metrics (OMs) and transforms them to Key Performance Indicators (KPIs) based on a industry-standard COBIT\textregistered{} framework. KPIs allow service providers to be able to capture the health of an underlying IT environment. With multiple use cases of KPIs, we demonstrated that our framework provides immense value to the Technical Architects not just in terms of monitoring the IT environment, but also understanding the actual causes of the problems that occur in the environment via KPI correlation analysis and KPI score prediction. Finally, we discuss the challenges and opportunities that Hybrid/Multi Cloud brings from the perspective of Cloud Service providers, by identifying challenges at IaaS and PaaS level, coupled with changes in traditional service management processes when applied to hybrid Cloud environments.

\newcommand{\BIBdecl}{\setlength{\itemsep}{0.25 em}}
\bibliographystyle{IEEEtran}
\bibliography{paper}

\end{document}